\documentclass[a4paper,12pt]{article}
\usepackage[dvips]{epsfig}

\advance\textheight2.8cm        
\advance\topmargin-2.0cm
\advance\textwidth2.4cm
\advance\evensidemargin-1.6cm   
\advance\oddsidemargin-1.6cm

\def\phi{\varphi}
\def\eps{\varepsilon}

\def\at#1{}

\begin{document}


\begin{center}

{\LARGE \bf A multi-phase, multi-component} \\~

{\LARGE \bf  critical equation of state} \\

\vspace{1cm}

\centerline{\sl {\large \bf Arnold Neumaier}}

\vspace{0.5cm}

\centerline{\sl Fakult\"at f\"ur Mathematik, Universit\"at Wien}
\centerline{\sl Nordbergstr. 15, A-1090 Wien, Austria}
\centerline{\sl email: Arnold.Neumaier@univie.ac.at}
\centerline{\sl WWW: \url{http://www.mat.univie.ac.at/~neum/}}

\vspace{0.5cm}


\end{center}

\hfill July 31, 2013

\vfill


\vfill

{\bf Abstract.} 
Realistic equations of state valid in the whole state space of a 
multi-component mixture should satisfy at least three important 
constraints:\\
(i) The Gibbs phase rule holds.\\
(ii) At low densities, one can deduce a virial equation of 
state with the correct multi-component structure.\\
(iii) Close to critical points, plait points, and consolute points, the 
correct universality and scaling behavior is guaranteed.

This paper discusses semiempirical equations of state for mixtures that 
express the pressure as an explicit function of temperature and the 
chemical potentials.
In the first part, expressions are derived for the most important 
thermodynamic quantities. The main result of the second part is the 
construction of a large family of equations of state with the 
properties (i)--(iii).


\newpage
\tableofcontents 

\vspace*{1cm}

\section{Activity equations of state}

We consider mixtures with $C$ pure \bfi{components} $i=1,\dots,C$.
Their equilibrium is characterized by constant \bfi{temperature} $T$, 
\bfi{pressure} $P$, and \bfi{chemical potentials} $\mu_i$ 
($i=1,\dots,C$), whereas the various thermodynamic densities and 
concentrations are piecewise constant only, with jumps at 
\bfi{phase boundaries}. We write
$n_i$ for the \bfi{mole number} of component $i$ and $V$ for the 
\bfi{total volume} of the mixture; in each phase, the \bfi{density} 
$n_i/V$ of component $i$ is determined by temperature, pressure, and 
chemical potentials. $R$ denotes the universal gas constant.
The theory of \bfi{thermodynamics} describes and derives the rules 
that relate these quantities; the axioms of thermodynamics (see
Appendix \ref{s.axioms}) itself can be derived by means of 
\bfi{statistical mechanics} from the dynamics of the molecules making 
up the mixture. For a rigorous and self-contained derivation see, e.g., 
\sca{Neumaier} \cite[Part II]{Neu.QML1}.

It is well-known (see, e.g., \sca{Callen} \cite{Cal}) that in the zero 
density limit, a mixture behaves like an ideal gas, characterized 
by chemical potentials\footnote{\label{f.ambig}
The thermodynamic formalism specifies each chemical potential $\mu_i$ 
only up to a term $\mu_{Ui}+T\mu_{Si}$ with arbitrary constants 
$\mu_{Ui}$ and $\mu_{Si}$. These constants depend on the values of 
internal energy $U$ and entropy $S$ (where only differences are 
observable) at some fixed reference point. The $\gamma_i(T)$ inherit
this ambiguity as the combined expression $\mu_i/RT-\gamma_i(T)$ 
is observable and hence unambiguous.
} 
of the form
\[
\mu_i=RT\Big(\gamma_i(T)+\log\frac{V_0n_i}{V}\Big)~~~ 
                                             \for \frac{V_0n_i}{V}\to 0,
\]
with a fixed reference volume $V_0>0$ and suitable 
functions $\gamma_i$ of $T$, and an equation of state of the form
\[
\frac{PV}{RT}=n_1+\ldots+n_C.
\]
We shall work with dimensionless reduced quantities, using a fixed 
reference temperature $T_0>0$ and a fixed reference pressure $P_0>0$
satisfying
\[
P_0V_0=RT_0,
\]
and formulate our results in terms of\footnote{
To aid the reader, we use $:=$ to indicate that a term on the left 
is defined by the right hand side. 
} 
the \bfi{reduced temperature}\footnote{
This choice, while somewhat unconventional, gives the simplest formulas
in Section \ref{s.free}, and was used already by 
\sca{Vidler \& Tennyson} \cite{VidT} for highly accurate fits to ideal 
gas properties of water.
} 
\[
\tau:=\log\frac{T}{T_0},
\]
the temperature-scaled \bfi{reduced pressure}
\[
\pi:=\frac{PT_0}{P_0T},
\]
the \bfi{reduced activities}
\[
z_i:=\exp\Big(\frac{\mu_i}{RT}-\gamma_i(T)\Big),
\]
which we collect together in a $C$-dimensional \bfi{activity vector} 
$z$, and the \bfi{reduced densities}
\lbeq{e.rhoi}
\rho_i:=\frac{V_0}{V}n_i~~~(i=1,\ldots,C).
\eeq
It is also conventional to use the \bfi{total molar number} 
\[
n:=\sum_i n_i
\]
and the \bfi{mole fractions}
\[
x_i:=\frac{n_i}{n}=\frac{\rho_i}{\rho}~~~(i=1,\ldots,C),
\]
where 
 \[
\rho=\sum_i\rho_i
\]
is the \bfi{total reduced density}. Clearly
\[
\sum_i x_i=1,~~~~~x_i\ge0~~(i=1,\ldots,C).
\]
The original variables can be easily recovered from the reduced 
quantities and the total volume $V$:
\lbeq{e.orig}
T=T_0e^\tau,~~~P=\frac{P_0T}{T_0}\pi,~~~
\mu_i=RT(c_i(\tau)+\log z_i),~~~n_i=\frac{V}{V_0}\rho_i,
\eeq
where
\[
c_i(\tau):=\gamma_i(T_0e^\tau)=\gamma_i(T).
\]
The ideal gas law takes the simple form
\lbeq{e.ideal}
\pi=\sum_i z_i, ~~~~~\rho_i=z_i~~(i=1,\ldots,C).
\eeq
In a general mixture, the intrinsic meaning of $\pi,\tau,z$, visible 
only in non-equilibrium situations, is that of \bfi{thermodynamic force 
fields} with a distinguished meaning whose spatial gradients induce 
changes in various densities. 
Equilibrium is characterized by the absence of these forces, i.e., the 
constancy of the corresponding force fields throughout the mixture.
$\tau$, $\pi$ and the $z_i$ are nonnegative, and
a (reduced) activity $z_i$ is zero (corresponding to $\mu_i=-\infty$) 
precisely when a substance is not present. In particular, pure 
substances are the limiting case of mixtures where only a single 
component of $z$ is nonzero. Intuitively, the reduced activities 
are a kind of effective densities (the name derives from the now 
outdated terminology ''active number densities''), but with a force 
field character.

Since the thermodynamic state space is only $(C+1)$-dimensional, there 
is always an algebraic relation between $\pi$, $\tau$, and $z$ 
generalizing \gzit{e.ideal}, which we shall call an
\bfi{activity equation of state} (activity \bfi{EOS}). 
In some sense, an activity EOS is the most fundamental formulation of 
equilibrium thermodynamics. Indeed, everything else in thermodynamics 
can be derived by assuming a few properties of an activity EOS only;
see Section \ref{s.free}.
Many phenomena find their simplest description when formulated in terms 
of the variables $\tau$, $\pi$, and $z$; cf. the discussion in 
\sca{Fisher} \cite{Fis0}.\footnote{
The following is taken from the text after (2.2): 
``\em{[...] the "forces" or "fields" rather than the "densities", are 
the more basic and simpler variables in terms of which to describe a 
statistical system undergoing a phase transition. 
The validity of this general philosophy is confirmed by nearly all 
solved problems in statistical mechanics; 
the greater simplicity of the grand canonical formulation for so
many purposes, particularly for the derivation of fluctuation formulae  
or sum rules, is a familiar illustration of the point. 
A further explicit demonstration will be provided by the soluble models
of systems displaying critical behavior [...] 
At a fundamental level the preferred status of the forces seems to 
correspond rather directly to the completely convex character of the 
corresponding thermodynamic potentials, alluded to above. 
Convexity in all the forces together means in effect that the various 
local densities can fluctuate freely with, at fixed fields, no over-all 
constraints.}''
} 

In statistical mechanics (see, e.g., \sca{Reichl} \cite{Rei}), one 
derives from the grand canonical ensemble in the thermodynamic limit an 
activity EOS of the pressure-explicit form
\lbeq{e.activityEOS}
\pi=\pi(\tau,z),
\eeq
and proves that the pressure can be written for small $z$ as a power 
series in $z$,
\lbeq{e.pi}
\pi(\tau,z)=\sum_{k_1,\ldots,k_C=0}^\infty 
                 \pi_{k_1,\ldots,k_C}(\tau)z_1^{k_1}\ldots z_C^{k_C}
\eeq
with smooth functions $\pi_{k_1,\ldots,k_C}(\tau)$ of $\tau$ that are 
(at least in principle) computable from microscopic information.
The way we defined the reduced variables implies that for small $z$ we 
have $\pi(\tau,z)=\sum z_i+O(z^2)$, hence
$\pi_{k_1,\ldots,k_C}(\tau)=0$ if all $k_i$ vanish, and 
$\pi_{k_1,\ldots,k_C}(\tau)=0$ if all $k_i$ but one vanish and the 
remaining takes the value $1$.

Therefore we shall take an equation with these properties as our 
starting point. However, we do not assume that $\pi(\tau,z)$ is derived 
from statistical mechanics but take it as an appropriate semiempirical 
expression whose form must be determined from constraints from a 
combination of theoretical considerations and matching to experimental 
data.
The main goal of this paper is to propose in Section \ref{s.multiphase} 
a particular class of activity equations of state with the properties 
stated in the abstract. EOSs from this class therefore have the 
potential to cover the whole state space. However, to relate the 
activity EOS \gzit{e.activityEOS} to more well-known forms of the EOS,
we first discuss in Sections \ref{s.free}--\ref{s.virial} more or 
less well-known consequences of \gzit{e.activityEOS} and \gzit{e.pi}, 
rederiving the most important formulas in a concise way.
Section \ref{s.phaseEq} discusses the additional issues introduced by
multiple phases.

\section{Free energies}\label{s.free}

In this section, we derive from the activity EOS formulas for mole 
numbers, entropy, various free energies, and their reduced versions.
At a given volume $V$, these and everything else may be computed from 
the expression
\[
P(T,\mu):=\frac{P_0T}{T_0}\pi(\tau,z),~~~~
\tau=\log\frac{T}{T_0},~~~
z_i=\exp\Big(\frac{\mu_i}{RT}-c_i(\tau)\Big),
\]
for $P$ at fixed $T$ and $\mu$, deduced from \gzit{e.orig}.
An activity EOS is thermodynamically consistent whenever $P(T,\mu)$
is jointly convex in $(T,\mu)$ and a strictly increasing function of 
each $\mu_i$ when the other variables are held constant; 
cf. Appendix \ref{s.axioms}. 
\at{Within a single phase, $P(T,\mu)$ is three times continuously 
differentiable; then a necessary and sufficient condition for convexity 
is that $P(T,\mu)$ has a positive semidefinite Hessian matrix.}

Fixing $T$ and all but one of the variables $\mu_1,\ldots,\mu_C$ in the 
\bfi{Gibbs--Duhem relation}
\lbeq{e.GibbsDuhem}
VdP=SdT+\sum_i n_i d\mu_i,
\eeq
where $S$ denotes the \bfi{entropy}, gives
\lbeq{e.ni}
\frac{dP(T,\mu)}{d\mu_i}=\frac{n_i}{V}=\frac{\rho_i}{V_0},
\eeq 
\[
\rho_i=\frac{V_0}{V}n_i=V_0\frac{dP(T,\mu)}{d\mu_i}
=\frac{V_0P_0T}{T_0}\frac{d\pi}{d\mu_i}
=RT\frac{d\pi}{dz_i}\frac{dz_i}{d\mu_i}
=z_i\frac{d\pi}{dz_i}.
\]
We conclude that
\lbeq{e.rhoi2}
\rho_i=z_i\frac{d\pi(\tau,z)}{dz_i} \for i=1,\ldots,C.
\eeq
Since the $\rho_i$ are nonnegative, $\pi(\tau,z)$ must be a strictly 
increasing function of each $z_i$ when the remaining variables are held 
constant. 
We describe the \bfi{Gibbs free energy}
\[
G:=\sum_i \mu_i n_i=\frac{RTV}{V_0}g
\]
and the \bfi{Helmholtz free energy}
\[
A:=G-PV=\frac{RTV}{V_0}f
\]
in terms of the dimensionless \bfi{reduced Gibbs free energy} $g$ and
the dimensionless \bfi{reduced Helmholtz free energy} $f$. A simple
substitution gives
\lbeq{e.gFromz}
g=\sum_i \rho_i(c_i+\log z_i),~~~f=g-\pi.
\eeq
The \bfi{internal energy}
\[
U:=A+TS=\frac{RTV}{V_0}\eps
\]
and the \bfi{enthalpy}
\[
H:=G+TS=\frac{RTV}{V_0}h
\]
are described in terms of the dimensionless \bfi{reduced internal 
energy} $\eps$, and the dimensionless \bfi{reduced enthalpy} $h$. 
Substitution now gives
\lbeq{e.allFromz}
\eps=f+s,~~h=g+s
\eeq
with the dimensionless \bfi{reduced entropy} $s=\D\frac{V_0S}{RV}$, 
which implies
\lbeq{e.free}
S=\frac{RV}{V_0}s.
\eeq
Fixing all variables $\mu_1,\ldots,\mu_C$ in \gzit{e.GibbsDuhem} gives
\lbeq{e.s}
\frac{dP(T,\mu)}{dT}=\frac{S}{V}=\frac{Rs}{V_0},
\eeq
\[
\bary{lll}
s&=&\D\frac{V_0S}{RV}=\frac{V_0}{R}\,\frac{dP(T,\mu)}{dT}
= \frac{V_0P_0}{RT_0}\,\frac{d(T\pi)}{dT}=\frac{d(T\pi)}{dT}
= \pi+T\D\frac{d\pi}{dT}\\[5mm]
&=&\pi+T\D\Big(\frac{d\pi}{d\tau}\frac{d\tau}{dT}
          +\sum_i \frac{d\pi}{dz_i}\frac{dz_i}{dT}\Big)
=\pi+\frac{d\pi}{d\tau}
-\sum_i \frac{d\pi}{dz_i}z_i\Big(\frac{dc_i}{d\tau}+c_i+\log z_i\Big)
\eary
\]
since
\[
\frac{dz_i}{dT}
=z_i\Big(-\frac{\mu_i}{RT^2}-\frac{dc_i}{d\tau}\,\frac{1}{T}\Big)
=-\frac{z_i}{T}\Big(\frac{dc_i}{d\tau}+\frac{\mu_i}{RT}\Big)
=-\frac{z_i}{T}\Big(\frac{dc_i}{d\tau}+c_i+\log z_i\Big).
\]
Therefore
\lbeq{e.spi}
s=\pi+\frac{d\pi}{d\tau}
-\sum_i \rho_i\Big(\frac{dc_i}{d\tau}+c_i+\log z_i\Big),
\eeq
\lbeq{e.hpi}
h=g+s=\pi+\frac{d\pi}{d\tau}-\sum_i \rho_i\frac{dc_i}{d\tau},
\eeq
and since $\eps=f+s=g-\pi+s=h-\pi$, we find that 
\lbeq{e.eps}
\eps=\frac{d\pi(\tau,z)}{d\tau}-\sum_i\rho_i\frac{dc_i(\tau)}{d\tau}.
\eeq
Looking at \gzit{e.orig}, \gzit{e.rhoi2}, \gzit{e.gFromz}, \gzit{e.spi},
\gzit{e.hpi}, and \gzit{e.eps}, we see that the activity EOS 
\gzit{e.activityEOS} encodes all thermodynamic information about the 
mixture. In particular, one can also compute thermodynamic response 
functions such as the (volume-based) \bfi{molar heat capacity}
\[
c_V(T):=\Big(\frac{dU}{n\,dT}\Big)_{V,n} 
= \frac{R}{\rho}\,\Big(\frac{d(T\eps)}{dT}\Big)_\rho
= \frac{R}{\rho}\Big(\eps + \frac{d\eps}{d\tau}\Big)_\rho.
\]
\at{add formulas for other important second-order thermodynamic 
quantities. \\
$c_P(T)=\Big(\frac{dH}{ndT}\Big)_{P,n}$ is less easy to evaluate as it 
needs $dV/dT$ at fixed $n$.}

In engineering practice, one usually works in terms of densities or 
composition rather than activities. To determine 
$z=z(\tau,\rho_1,\ldots,\rho_C)$ as a function of reduced temperature 
and densities, one must solve the nonlinear system of equations 
\gzit{e.rhoi2}.
\at{Does monotonicity imply unique solvability for $z$ in terms of 
$\rho$?}
Often, \gzit{e.rhoi2} is uniquely solvable for the $\rho_i$, but if 
there are multiple solutions, the solution with largest reduced
Helmholtz free energy $f$ is the correct one.
\at{growth constraints imposed by boundedness of energy?}
If we substitute this into the activity EOS \gzit{e.activityEOS},
we find an EOS 
\lbeq{e.pirho}
\pi=\pi\Big(\tau,z(\tau,\rho_1,\ldots,\rho_C)\Big)
\eeq
for the reduced pressure in terms of reduced temperature and densities.
We also find (with $\tau=\log(T/T_0)$) the isochoric form
\lbeq{e.frho}
f(T,\rho_1,\ldots,\rho_C)
=\sum_i \rho_i\Big(c_i(\tau)+\log z_i(\tau,\rho_1,\ldots,\rho_C)\Big)
-\pi\Big(\tau,z(\tau,\rho_1,\ldots,\rho_C)\Big)
\eeq
of the reduced Helmholtz free energy as discussed, e.g., by 
\sca{Sengers \& Levelt Sengers} \cite{SenL} and 
\sca{Qui\~nones-Cisneros \& Deiters} \cite{QuiD}.

\section{Ideal gas and virial equation of state}
\label{s.virial}

Specializing the preceding formulas to the case of a pure ideal gas with
the single component $i$, we find
\[
\pi=z_i=\rho_i,
\]
\[
g=\rho_i(c_i+\log \rho_i),~~~
f=g-\pi=\rho_i(c_i-1+\log \rho_i),
\]
\[
\eps=-\rho_i\frac{dc_i}{d\tau},~~~
h=\rho_i\Big(1-\frac{dc_i}{d\tau}\Big),~~~
s=h-g=\rho_i\Big(1-\frac{dc_i}{d\tau}-c_i-\log \rho_i\Big).
\]
\[
c_V(T)=-RT\frac{d^2(Tc_i)}{dT^2}
=-R\Big(\frac{dc_i}{d\tau}+\frac{d^2c_i}{d\tau^2}\Big).
\]
In particular, one can calculate $c_i(\tau)$ from zero density heat 
capacities by numerical integration.\footnote{
The different ways of fixing the ambiguities in the definition of the 
$\mu_i$ mentioned in footnote ${}^{\ref{f.ambig}}$ correspond to 
the different possible choices of the integration constants.
} 
Real gases behave like an ideal gas only in the limit of zero densities.
At low densities, the behavior of mixtures is governed 
by the multi-component \bfi{virial equation of state}, an expansion of 
the pressure as a multivariate power series in densities with 
temperature-dependent coefficient, first derived by \sca{Fuchs} 
\cite{Fuc} in statistical mechanics terms.
But the virial expansion may be obtained from \gzit{e.pirho} without 
reference to statistical mechanics by a multivariate Taylor expansion. 
Given an activity EOS \gzit{e.activityEOS} satisfying \gzit{e.pi} and 
reducing to $\pi=\sum z_i$ in the zero density limit, we may write 
\gzit{e.pi} at low reduced densities $\rho_i$ as
\[
\pi=\sum_i z_i + \half \sum_{i,j} G_{ij}z_iz_j +O(z^3),
\]
leaving implicit the temperature dependence of the coefficients
\[
G_{ij}(\tau)=\frac{d^2\pi}{dz_idz_j}(\tau,0)
\]
that determine the second order deviations from the ideal gas law.
At low densities, the system of equations 
\gzit{e.rhoi2} for $i=1,\ldots,C$ can be solved uniquely for 
$z_i=\rho_i+O(\rho^2)$ by successive substitution. Since 
\[
z_i=\frac{\rho_i}{d\pi/dz_i}
=\frac{\rho_i}{\D 1+\sum_j G_{ij}z_j+O(z^2)}
=\rho_i\Big(1-\sum_j G_{ij}\rho_j+O(\rho^2)\Big),
\]
we have
\[
\bary{lll}
\pi&=&\D\sum_i z_i + \half \sum_{i,j} G_{ij}z_iz_j +O(z^3)\\[5mm]
&=&\D
\sum_i \rho_i\Big(1-\sum_j G_{ij}\rho_j+O(\rho^2)\Big)
     + \half \sum_{i,j} G_{ij}\rho_i\rho_j +O(\rho^3)\\[5mm]
&=&\D\sum_i \rho_i-\half \sum_{i,j}G_{ij}\rho_i\rho_j
+O(\rho^3).
\eary
\]
This gives to second order the \bfi{reduced virial equation of state}  
\lbeq{e.virial}
\pi=\sum_i \rho_i-\half\sum_{i,j}G_{ij}(\tau)\rho_i\rho_j+O(\rho^3)
=\rho-\frac{\rho^2}{2}\sum_{i,j}G_{ij}(\tau)x_ix_j+O(\rho^3).
\eeq
Using \gzit{e.orig}, we may compare this with the traditional, 
unreduced form
\[
\frac{P}{RT} = \frac{n}{V} + B_2(T,x) \Big(\frac{n}{V}\Big)^2 
                                   + O\Big(\Big(\frac{n}{V}\Big)^3\Big)
\]
defining the \bfi{second virial coefficient} $B_2(T,x)$. We find that
\[
B_2(T,x)
=-\frac{T_0R}{2P_0}\sum_{i,j}G_{ij}\Big(\log\frac{T_0}{T}\Big)x_ix_j.
\]
Higher order terms in the virial equation of state may be derived in a 
similar fashion.

\at{Work out $g$, $\eps$, etc. to second order.
Deduce the value of the \bfi{activity coefficients} $\gamma_i$ defined 
by $RT\log\gamma_i=dG^E(T,p,n)/dn_i$.}

\section{Phase equilibrium}
\label{s.phaseEq}

In a mixture that exists in multiple phases, $\pi(\tau,z)$ is 
nondifferentiable along the \bfi{coexistence manifold}, defined as 
the set of states $(\tau,z)$ for which arbitrarily small neighborhoods 
contain points corresponding to two different phases. Each phase $s$ 
has its own three times\footnote{
In theory, we have infinite differentiability away from the coexistence 
manifold. However for practical modeling, it is sufficient to require
three times continuous differentiability. Then all thermodynamic 
quantities of interest, being obtainable from $\pi(\tau,z)$ and its 
first two derivatives, have a continuously differentiable dependence 
on $\tau$ and $z$, which is sufficiently smooth for practical 
applications.
} 
continuously differentiable activity EOS 
\[
\pi=\pi_s(\tau,z).
\]
In each single phase region, only one of these equations holds, while 
on the coexistence manifold, two or more of these equations are 
satisfied simultaneously. For an $m$-phase equilibrium involving 
the phases $s_1,\ldots,s_m$, we have $m$ independent equations
\[
\pi=\pi_s(\tau,z)~~~(s=s_1,\ldots,s_m).
\]
Generically, these define a $(C+2-m)$-dimensional manifold, which is 
the well-known \bfi{Gibbs phase rule}. In particular, the number of 
coexistent phases is generically at most $C+2$.

Rigorous results from statistical mechanics (cf. \sca{Langer} 
\cite{Lan}, \sca{Penrose \& Lebowitz} \cite{PenL}, 
\sca{Isakov} \cite{Isa}, \sca{Friedli \& Pfister} \cite{FriP}) 
imply that any EOS has essential singularities\footnote{
An essential singularity is a point at which the Taylor series 
expansion has zero convergence radius.
} 
everywhere along the 
coexistence manifold.
These are absent only in an approximate mean field treatment and in 
semiempirical models, where each $\pi_s(\tau,z)$ is usually three 
continuously differentiable in some region beyond the coexistence 
boundaries, until singularities are reached at another boundary 
defining the \bfi{spinodal manifold}. In the region between the 
coexistence and the spinodal manifold, the phase activity EOS then 
describes a metastable state. These states are strictly speaking not 
governed by equilibrium thermodynamics, but as long as the relaxation 
times to the stable state are sufficiently long, an approximate 
equilibrium description is possible, and it is accurate close to the
coexistence manifold. In any case, the stable state is the state of 
maximal pressure\footnote{
For example, consider a $(P,\mu)$ diagram for two phases of a pure 
substance at fixed temperature. Here $\mu$ is essentially the Gibbs 
free energy; therefore the stable phase consists of the branches with
the smallest value of $\mu$ at fixed $P$. The drawing then implies that 
$P$ has the largest value at fixed $\mu$.
} 
and satisfies the \bfi{multiphase activity EOS}
\lbeq{e.phaseEOS}
\pi=\pi(\tau,z):=\max_s \pi_s(\tau,z),
\eeq  
where at each $(\tau,z)$ the maximum is taken over all possible phases 
$s$, using the value $-\infty$ at states beyond the spinodal where 
some $\pi_s(\tau,z)$ is undefined.
\at{Note that the maximum of convex functions is convex!}

Modeling phases by means of \gzit{e.phaseEOS} has the advantage 
compared to models based on a Helmholtz free energy that the model
cannot have any accidental unwanted phases. Coexistent phases are
characterized by common values of $\pi$, $\tau$, and $z$. In particular,
stable coexistent phases exist precisely at the states in which several
of the functions $\pi_s(\tau,z)$ have coinciding values. 

To derive the composition of each phase at fixed $\pi$, $\tau$, and
total composition $x$ in stable equilibrium, one must solve the 
constrained optimization problem\footnote{
The formulation as a maximization problem is due to thermodynamic 
stability; see Appendix \ref{s.axioms}. It is similar to the 
maximization occuring in the definition of Legendre transforms and in 
the definition of the Gibbs potential in \sca{Neumaier} 
\cite{Neu.phenTherm}.
} 
\[
\bary{ll}
\D\max_z & \D\sum_i x_i\log z_i\\
\st  & \pi_s(\tau,z)\le \pi ~~~\mbox{for all possible phases $s$}.
\eary
\]
By general results from optimization theory, any solution of this 
optimization problem must satisfy the \bfi{optimality conditions} 
\lbeq{e.opti}
x_i=\sum_s \lambda_s\rho_{si},~~~
\rho_{si}=z_i\frac{d\pi_s(\tau,z)}{dz_i},
\eeq
where the $\lambda_s$ are Lagrange multipliers satisfying the
\bfi{complementarity conditions}
\lbeq{e.comp}
\min(\lambda_s,\pi-\pi_s(\tau,z))=0~~~
                                     \mbox{for all possible phases $s$}.
\eeq
Should the optimality conditions have multiple solutions, the correct
solution is given by the global maximum.
The reduced density $\rho_s$ of phase $s$ and the mole fractions 
$x_{si}$ of component $i$ in phase $s$ are then given by
\[
\rho_s=\sum_i \rho_{si},~~~
x_{si}=\frac{\rho_{si}}{\rho_s}.
\]
Compared with the optimization problems arising in the conventional 
Helmholtz or Gibbs formulations, there are two important differences:\\
(i) A solution of the optimization problem is automatically a stable 
equilibrium; to obtain a metastable phase equilibrium one must solve
instead an alternative optimization problem in which the constraints 
corresponding to the more stable competing phases are dropped.\\
(ii) The complementarity conditions automatically determine the phases
actually present as they imply $\lambda_s=0$ if $\pi_s(\tau,z)<\pi$.

To generalize this to reacting mixtures, we consider an initial 
composition and the final equilibrium state resulting if the mixture is 
left to reach both chemical and phase equilibrium. As any chemical 
reaction changes the mole numbers of some of the components, the 
composition and the total number of moles changes until equilibrium 
is reached. The phase equilibrium problem in the presence of chemical 
reactions is therefore obtained by solving the constrained optimization 
problem
\[
\bary{ll}
\D\max_{\xi,z} & \D\sum_i \Big(n_i^\init+(N\xi)_i\Big)\log z_i\\
\st  & \pi_s(\tau,z)\le \pi ~~~\mbox{for all phases $s$},\\
     &  0\le n_i^\init+(N\xi)_i~~~(i=1\dots,C).
\eary
\]
Here $n_i^\init$ is the initial mole number of component $i$, $\xi$ is 
a vector of \bfi{extent of reaction} coordinates, and 
$n_i=n_i^\init+(N\xi)_i$ the equilibrium mole number of component $i$. 
$N$ is the matrix of stoichiometric coefficients of the allowed 
chemical reactions, each column of $N$ corresponding to one chemical 
reaction. 
\at{Or use $Mx=Mx^\init$, where each row of $M$ corresponds to atoms or 
chemical groups left invariant by the allowed chemical reactions.}

Traditionally, the semiempirical view is microscopically supported 
by an approximate mean field treatment in terms of an analytic 
Helmholtz free energy. In this case the state functions in different 
phases are connected by analytic continuation using so-called van der 
Waals loops, which contain besides the metastable regions also 
spurious unphysical states. As a consequence of the van der Waals 
construction, the $\pi_s(\tau,z)$ are located on different sheets of 
the same analytic function. However, when presented in the form of an 
activity EOS \gzit{e.phaseEOS}, the $\pi_s(\tau,z)$ for different 
phases of the same mixture may be completely unrelated. 
Such a heterogenous approach is common practice for the joint modeling 
of fluid and solid phases, though there are frameworks in which one
can give a joint Helmholtz description of all phases; see, e.g., 
\sca{Lomonosov} \cite{Lom}.

However, if some of the phases are not well separated everywhere in the 
state space then a more sophisticated joint description must exist. 
This is the case precisely when a coexistence boundary contains a 
\bfi{critical point}, i.e., a point on the boundary of the coexistence 
manifold (\sca{Griffiths \& Wheeler} \cite{GriW}). In this case, the
joint description is constrained by nontrivial scaling laws near the 
critical points, which will be discussed next.

\section{A multiphase critical equation of state}\label{s.multiphase} 

Equations of state valid over the whole state space must account 
for the existence of critical points (for pure fluids a liquid-vapor 
critical point, for mixtures plait points at liquid-vapor equilibrium 
and consolute points at liquid-liquid equilibrium) and of the 
\bfi{universal}, i.e., substance independent power laws with which 
certain thermodynamic quantities scale near the critical point; 
cf. the fairly recent survey by \sca{Sengers \& Shanks} \cite{SenS}.

The form of any globally valid equation of state (EOS) is strongly 
restricted by renormalization group arguments. An exposition of the
results in a form most useful for multi-component mixtures is given in
\sca{Neumaier} \cite{Neu.critScal}, where a new and general analytic 
form of an EOS accomodating all critical properties was drived.
In this setting, the critical behavior of mixtures is 
characterized by the existence of several \bfi {scaling fields}, 
the \bfi{strong scaling field} $\Sigma$ the \bfi{thermal scaling field} 
$\Theta$, the \bfi{dependent scaling field} $D$, and a number of 
\bfi{scaling correction fields} $I_k$ ($k=1,2,\ldots$); here the term 
\bfi{field} just means a smooth function of\footnote{
Since the renormalization group technique works rigorously in terms of 
thermodynamic force fields only, the variables $\tau$, $\pi$, and $z$ 
are the most natural thermodynamic variables for a multi-component 
equation of state near the critical point.
} 
$\pi$, $\tau$, $z$. The form of the analytic scaling EOS is
\lbeq{e.sig2}
D^{-2a/e}\Sigma^2
=\Gamma(D^{-1/e}\Theta,D^{-e_1/e}I_1,D^{-e_2/e}I_2,\ldots),
\eeq
where $\Gamma$ is a universal, substance-independent analytic function 
of its arguments; the only nonanalytic behavior is in the powers of $D$,
involving the \bfi{critical exponents}
\lbeq{e.critEx}
e>a>1>0>e_1 \ge e_2 \ge \ldots .
\eeq
The critical exponents are also universal; the first few values are 
approximately given by
\lbeq{e.cd}
a \approx 1.56383(34),~~
e \approx 1.89036(48),~ 
e_1 \approx -0.52(3),~
e_2\approx -1.05(8),~
e_3\approx -1.5(3).
\eeq
Much is known about the universal function $\Gamma$; see 
\cite{Neu.critScal}.

The strong scaling field $\Sigma$ and the thermal scaling field $\Theta$
have a clear physical meaning:
For $\Theta\le 0$, the system is in a lower density phase if $\Sigma>0$
and in a higher density phase if $\Sigma<0$. 
If $\Sigma =0$, the system has two coexistent lower and higher density 
phases.  
At $\Theta<0=\Sigma$, we have a \bfi{first-order phase transition}, 
i.e., some thermodynamic response functions possess a jump 
discontinuity.
The inequality $\Theta>0$ defines a low density part of the phase space,
connected very smoothly to the low density phase at $\Sigma>0>\Theta$.
However, at $\Theta=0>\Sigma$, there is a \bfi{higher-order phase 
transition} between high density states and low density states, with 
continuous but nonanalytic thermodynamic response functions.
Near the vapor-liquid critical point, the condition $\Theta>0>\Sigma$ 
corresponds approximately to the conventional definition of 
supercritical, which in an engineering context means that both pressure 
and temperature are above the pressure and temperature at the critical 
point. This justifies to call the states with $\Theta>0>\Sigma$ 
\bfi{supercritical} in a theoretically more justified sense.
Critical points are characterized by $\Sigma =\Theta=0$. 
Generically, for mixtures with $C$ components, they form a manifold of 
dimension $C-1$ in the $(C+1)$-dimensional thermodynamic state space.

The above is referred to as the \bfi{complete scaling} setting.
Much of the work on critical EOS has been done in the more restrictive, 
simplified setting of \bfi{revised scaling}. The latter means that $D$ 
has the special form
\[
D(\pi,\tau,z)=\pi_\regu(\tau,z)-\pi
\]
with a smooth function $\pi_\regu$ of $\tau$ and $z$, and that the 
remaining scaling fields $\Sigma$, $\Theta$ and the $I_k$ are 
independent of $\pi$. In the resulting activity EOS
\[
\pi=\pi_\regu(\tau,z)-\wt D(\tau,z),
\]
where in the present context $\wt D(\tau,z)$ is obtained from 
\gzit{e.sig2} by solving for $D$, one interprets $\pi_\regu(\tau,z)$
as a classical, regular part and $\wt D(\tau,z)$ as a singular 
\bfi{crossover term} to critical behavior.

In the first paper modeling (binary) mixtures with correct scaling 
properties close to the critical point, \sca{Leung \& Griffiths} 
\cite{LeuG} express everything in terms of force field variables.
For industrial applications, multicomponent formulations in a cubic 
EOS, Helmholtz or Gibbs free energy framework are desirable, so that 
the composition can be kept constant.
A number of such formulations were presented in the literature; see, 
e.g., \cite{AniGKS,BelKR,EdiAS,JinTS,KisE,KisF,KisR,RaiF}.
However, formulations at fixed composition cannot match exactly the
singularities; indeed \sca{Wheeler \& Griffiths} \cite{WheG}
prove that when some mole fractions are held constant, curves of plait 
points have bounded heat capacity, while the heat capacity at the 
critical point of a pure substance diverges.
A more detailed analysis leads to additional renormalization phenomena 
(\sca{Fisher} \cite{Fis0}). Ignoring these, as often done in these 
formulations, therefore requires additional approximations which may 
result in artifacts very close to the critical point 
(\sca{Kiselev \& Friend} \cite{KisF}). 
Other papers (e.g., \cite{CaiP,CaiQZH}) implement the renormalization 
group approach more directly, resulting in an iterative definition of an
equation of state that in the limit of infinitely many iterations 
satisfies the correct scaling laws.
A thorough discussion of many practically relevant issues is given in 
the surveys by \sca{Anisimov \& Sengers} \cite{AniS} and \sca{Behnejad}
et al. \cite{BehSA}. 

Most previous critical EOS are limited in several different ways:\\
$\bullet$
Frequently, the thermal scaling field $\Theta$ is taken to be linear in 
the temperature. However, linearity in $T$ limits the EOS to a narrow 
range of temperatures. Scaling fields with a more favorable temperature 
dependence such as one linear in $T^{-1}$ or $\tanh(T_0/T)$ 
extrapolate much better to the high temperature regime; see, e.g., 
{\sc Lundow \& Campbell} \cite{LunC}.\\
$\bullet$ 
Almost all studies attempting to go beyond the immediate neighborhood 
of the critical point work in the simplified setting of revised scaling.
However, this setting does not account for all observable fluid 
behavior; see, e.g., \sca{Bertrand} et al. \cite{BerNA}.
The only previous EOS not restricted to revised scaling is the 
crossover EOS of \sca{Bakhshandeh \& Behnejad} \cite{BakB}, which 
employs complete scaling, with scaling fields linear in $P$, $T$, and 
$\mu$.\\
$\bullet$ 
All noniterative equations of state with correct critical scaling are 
currently based on an implicit representation in terms of a Schofield 
type parameterization.\\ 
$\bullet$ 
Most papers discuss the two-phase case only. 
The only exception is \sca{Rainwater} \cite{Rai}, who attempts to cover 
vapor-liquid-liquid equilibrium. 
The main reason for this lack of generality seems to be that (as 
Rainwater's paper shows) a crossover mechanism in terms of a 
Schofield type parameterization is very difficult to extend to the 
multiphase case, since the Schofield parameters have no clear meaning 
far from the critical point and tend to introduce unphysical artifacts
such as two coexisting vapor phases. 

The new analytic scaling EOS \gzit{e.sig2} leads to a large family of 
implicit or explicit activity equations of state that, by their very 
form, automatically have the following properties:\\
(i) The Gibbs phase rule holds.\\
(ii) At low densities, one can deduce a virial equation of state with 
the correct multi-component structure.\\
(iii) Close to critical points, plait points, and consolute points, the 
correct universality and scaling behavior is guaranteed.

To find the most general form of an EOS with these properties, we 
partition the phases into groups $g$ completely separated by a phase 
space boundary, while the phases within each group may be connected 
with each other by paths in phase space not crossing the coexistence 
manifold. 
One of these groups consists of all fluid phases; for solid phases, the 
groups may consist of a single phase only or (as, e.g., for 
$\beta$-brass, cf. \sca{Lamers \& Schweika} \cite{LamS}) may contain 
several phases related by a critical point.
The phases within each group are described by common, 
substance-specific scaling fields $\Theta_g(\pi,\tau,z)$, 
$\Sigma_g(\pi,\tau,z)$, and $D_g(\pi,\tau,z)$, one for each phase 
group $g$. 
As discussed above, the critical behavior is correctly modelled if the 
EOS for each phase group takes the form
\lbeq{e.critEOS}
\Sigma_g^2=D_g^{2a/e}\Gamma(D_g^{-1/e}\Theta_g,
                              D_g^{e_1/e}I_{g1},D_g^{e_2/e}I_2,\ldots)
\eeq
with a universal, substance-independent function $\Gamma$; here the
arguments $(\pi,\tau,z)$ were suppressed for easy readability.
For given $\tau$ and $z$, the stable phase is determined by finding for 
each phase group $g$ the solutions $\pi$ of the equations 
\gzit{e.critEOS}, and taking  -- according to Section \ref{s.phaseEq} --
the one with largest $\pi$. In case of ties, multiple phases from 
different phase groups coexist.

Since the group $f$ of fluid phases contains the vapor phase, we 
require for this phase group additional properties that ensure a proper
ideal gas limit ($\pi$ and $z$ small). We require four conditions; the 
first condition makes sure that for small $\pi$ and $z$, we only have 
a single phase. The second condition allows us to write the EOS in
the pressure-explicit form \gzit{e.activityEOS} of an activity EOS.
The third condition embodies the ideal gas law, and the final condition 
produces the virial equation of state:
$\bullet$ 
\gzit{e.critEOS} holds identically for arguments 
$(\pi,\tau,z)=(0,\tau,0)$ with 
\lbeq{e.Wideal}
\Theta_f(0,\tau,0)\ge 1 \Forall \tau>0.
\eeq
This guarantees that $\Theta_f>0$ for small $\pi$ and $z$, so that we 
are in the single-phase case.\\
$\bullet$ 
The nondegeneracy condition
\[
\frac{dD_f(0,\tau,0)}{d\pi}\ne 0
\]
holds. This condition ensures that for small $\pi$ and $z$ we may 
uniquely  solve the equation $D_f(\pi,\tau,z)=D$ for 
$\pi=\pi_f(D,\tau,z)$.\\
$\bullet$ 
A condition on the derivative of \gzit{e.critEOS} at 
$(\pi,\tau,z)=(0,\tau,0)$ that guarantees that \gzit{e.critEOS} is 
consistent with (and hence implies) 
$\pi_f(D,\tau,z)=\sum_i z_i +O(z^2)$.\\
$\bullet$ 
$D_f$, $\Theta_f$, and $\Sigma_f$ have a multivariate power series 
expansion in $z$. 
This condition allow one to write $\pi_f(D,\tau,z)$ has a power series 
expansion in $z$. \\
As shown in Section \ref{s.virial}, these conditions together imply the 
virial equation of state and hence the correct low density behavior. 

Whenever all requirements discussed in this section are satisfied 
(which is easy to achieve), \gzit{e.critEOS} is a 
\bfi{global multi-component EOS} with correct critical scaling and a 
correct low density limit. 
For binary mixtures, a global crossover EOS with these properties was 
first derived by \sca{Kiselev \& Friend} \cite{KisF}, using the 
Schofield parameterization. The present approach is more general, needs 
no parameterization, and works for arbitrarily many components and 
phases. The fact that we use essentially the multiphase activity EOS 
\gzit{e.phaseEOS} from Section \ref{s.phaseEq} automatically ensures 
the Gibbs phase rule. 

Thus we satisfied the desired properties (i)--(iii).
The requirements discussed above still leave very much freedom for 
detailed modeling. For example, we may make the $\Theta_g$, 
and $\Sigma_g$ independent of $\pi$ and choose the $D_g$ linear in 
$\pi$, thus staying within the realm of the assumption of revised 
scaling. However, this would not be adequate for many real fluids;
so more general choices are advisable.

To model \bfi{vapor-liquid-liquid} equilibrium (VLLE), the simpler 
revised scaling setting is indeed not sufficient.
The vapor-liquid equilibrium is described by a coexistence 
equation relating $\tau$ and $z$, which we may write conceptually in 
the form $\tau=\tau_{VL}(z)$. The degree of freedom lost by enforcing
this coexistence relation reappears as an order parameter specifying 
the relative proportion of the vapor and liquid phases. Similarly,
liquid-liquid equilibrium is described by a coexistence equation 
$\tau=\tau_{LL}(z)$, and an order parameter specifying the relative 
proportion of the two liquid phases. When the two coexistence surfaces 
meet, i.e., if $\tau=\tau_{VL}(z)=\tau_{LL}(z)$, we have exchanged two 
lost degrees of freedom by two order parameters specifying the relative 
proportion of the vapor and the two liquid phases. 
Now both the VL and the LL branch of the coexistence manifold must 
satisfy the equation $\Sigma(\pi,\tau,z)=0$. This is possible only if 
$\Sigma$ is nonlinear; cf. \sca{Anisimov} et al. \cite{AniGKS}.
However, in revised scaling, $\Sigma$ is independent of $\pi$, and
$\pi$ is uniquely determined by $\tau$ and $z$ \cite{Neu.critScal}.
This implies that in revised scaling only two simultaneous phases 
within the same phase group are possible, which excludes VLLE.
However, already allowing $\Sigma(\pi,\tau,z)$ to be quadratic in $\pi$
removes this obstacle. 

For practical applications, it remains to be seen which particular
analytic choices for the scaling fields account for the behavior of 
mixtures in the whole state space. 
To be able to fit experimental data, one needs to choose particular 
forms for the scaling fields $\Theta$ and $\Sigma$ and the various 
coefficient functions. 
The strong scaling field $\Sigma_g$ of each group $g$ of phases is more 
or less determined by the requirement that it vanishes precisely on 
the coexistence curve, which allows it to be fitted directly to 
experimental coexistence data. 
Constraints for $\Sigma_f$, $\Theta_f$ and $D_f$ for the group $f$ 
of fluid phases are obtained by \gzit{e.Wideal} and by matching its 
$z=0$ expansion to the virial equation of state. 
Each thermal scaling field $\Theta_g$ is also constrained by the 
requirement that $\Theta_g$ vanishes on the critical manifold. 
Close to the critical manifold, the universal functions are constrained 
by the known asymptotic results, summarized in \cite{Neu.critScal}.
Universality gives further constraints when data for different mixtures 
are available, as the universal function $\Gamma$ must be independent 
of the particular mixture. 
The freedom remaining must be determined from experimental information. 
To exploit the available freedom without incurring artifacts due to 
excessive parameter sensitivity, fitting procedures may make use of all 
techniques available for the construction of modern, accurate 
multiparameter equations of state, as reviewed, e.g., in 
\sca{Lemmon \& Span} \cite{LemS}.

\appendix

\section{Axioms for equilibrium thermodynamics}
\label{s.axioms}

The theory of thermodynamics makes essential use of the concept 
of convexity. A set $X\subseteq \Rz^n$ is called \bfi{convex} if 
$tx+(1-t)y\in X$  for all $x,y\in X$ and all $t\in[0,1]$.
A real-valued function $\Phi$ is called \bfi{convex} on the convex set
$X\subseteq \Rz^n$ if $\Phi$ is defined on $X$ and, for all $x,y\in X$,
\[
\Phi(tx+(1-t)y) \le t\Phi(x)+(1-t)\Phi(y) \for 0\le t\le 1.
\]
If $x$ is written explicitly as several arguments (such as $T,P,\mu$ 
below), one says that $\Phi$ is \bfi{jointly convex} in these arguments.
Clearly, $\Phi$ is convex iff for all $x,y\in X$, the function 
$f:[0,1]\to \Rz$ defined by
\[
f(t):=\Phi(x+t(y-x))
\]
is convex. It is well-known that, for twice continuously 
differentiable $\Phi$, this is the case iff the second derivative 
$f''(t)$ is nonnegative for sufficiently small $t\ge 0$.

\sca{Neumaier} \cite{Neu.phenTherm} shows that the assumptions in the 
following list of axioms are sufficient to deduce all general results
of phenomenological thermodynamics of a single phase, including the 
extremal principles.

\begin{dfn}\bfi{(Phenomenological thermodynamics)}
\label{d.phen}\\
(i) Temperature $T$ and volume $V$ are positive, entropy $S$ and
mole numbers $n_i$ are nonnegative. 
The \bfi{extensive variables} $U,S,V,n_i$ are additive under the 
composition of disjoint subsystems. 

(ii) The \bfi{intensive variables} $T,P,\mu$ are related by the 
\bfi{equation of state}
\lbeq{e.def}
\Delta(T,P,\mu)=0.
\eeq
The \bfi{system function} $\Delta$ appearing in the equation of state 
is jointly convex in $T,P,\mu$ and decreasing in $P$. 
The set of $(T,P,\mu)$ satisfying $T>0$ and the equation of state is 
called the \bfi{state space}.

(iii) The internal energy $U$ satisfies the \bfi{Euler inequality}
\lbeq{e.Ui}
U\ge T S - P V + \sum_i \mu_i n_i
\eeq
for all $(T,P,\mu)$ in the state space.

(iv) \bfi{Equilibrium states} have well-defined intensive and extensive 
variables satisfying equality in \gzit{e.Ui}.
A system is in \bfi{equilibrium} if it is completely characterized by 
an equilibrium state.
\end{dfn}

It is proved in \cite{Neu.phenTherm} that, as a consequence, in 
any equilibrium state, the extensive variables are given by 
\lbeq{e.Sp}
S=\Omega\frac{\partial \Delta}{\partial T}(T,P,\mu),~~~
V=-\Omega\frac{\partial \Delta}{\partial P}(T,P,\mu),~~~
n_i=\Omega\frac{\partial \Delta}{\partial \mu_i}(T,P,\mu),
\eeq
and the \bfi{Euler equation}
\lbeq{e.Up}
U=T S - P V + \sum_i \mu_i n_i,
\eeq
the case of equality in \gzit{e.Ui}.
Here $\Omega$ is a positive number independent of $T$, $P$, and $\mu$,
called the \bfi{system size}. 

The system function $\Delta$ is not uniquely determined by a 
thermodynamic system, as multiplication by a nonzero function does not 
change the equation of state. Within a single phase $s$, this freedom 
may 
be used to bring the system equation into the pressure-explicit form
\[
\Delta(T,P,\mu)=P_s(T,\mu)-P
\]
with a suitable function $P_s(T,\mu)$.
In this case, the system size is the volume, $\Omega=V$, and the 
requirements that $P_s(T,\mu)$ is jointly convex in $(T,\mu)$ and 
\[
\frac{\partial P_s}{\partial \mu_i}(T,P,\mu)=\frac{n_i}{V}\ge 0
\]
are sufficient to satisfy the remaining conditions in the above axioms
with the equation of state 
\lbeq{e.PTmu}
P=P_s(T,\mu). 
\eeq
The equation of state for the most stable phases is given by 
$P=P(T,\mu)$ with the function
\[
P(T,\mu):=\max_s P_s(T,\mu),
\]
which is automatically convex when all $P_s(T,\mu)$ are convex.
The coexistence region consits of the states $(T,\mu)$ for which 
the maximum is attained for two or more phases $s$ simultaneously.

In terms of the reduced force fields $\pi$, $\tau$, and $z$, 
\gzit{e.PTmu} is just an activity EOS with
\lbeq{e.piFromDelta}
\pi(\tau,z):=\frac{\tau}{P_0}P_s(T,\mu),~~~~
T=\frac{T_0}{\tau},~~
\mu_i=\frac{RT_0}{\tau}\log\frac{z_i}{c_i(\tau)}.
\eeq
Conversely, one can get from any activity EOS \gzit{e.activityEOS}
an equation of state $P=P_s(T,\mu)$ with the definition
\lbeq{e.DeltaFrompi}
P_s(T,\mu):=\frac{P_0T}{T_0}\pi(\tau,z),~~~~
\tau=\frac{T_0}{T},~~z_i:=c_i(T_0/T)e^{\mu_i/RT},
\eeq
\at{note for multiple phases that the maximum of convex functions is 
convex.}

\bigskip
{\bf Acknowledgments.}
The author acknowledges with pleasure several discussions with
Ali Baharev on earlier versions of this manuscript, which lead to 
significant improvements.

\at{$M$ serves as an \bfi{order parameter}.
Because of approximations valid only very close to the critical point,  
the locus $\Sigma=0$ is called the \bfi{critical isochore}, and the 
locus $\Theta=0$ is called the \bfi{critical isotherm}.
But all this is irrelevant here.}

\at{uncited:
\sca{Griffiths} \cite{Gri},
\sca{Kiselev \& Ely} \cite{KisE.cr},
\sca{Mistura} \cite{Mis}.}
\

\bigskip

\end{document}